\documentclass[conference]{IEEEtran}
\usepackage{amsmath,epsfig,amssymb,verbatim,subfigure,color}
\usepackage{flushend}
\usepackage{cite}
\usepackage{lipsum}

\def\BibTeX{{\rm B\kern-.05em{\sc i\kern-.025em b}\kern-.08em
    T\kern-.1667em\lower.7ex\hbox{E}\kern-.125emX}}

\begin{document}

\title{Coverage Analysis of Relay Assisted Millimeter Wave Cellular Networks with Spatial Correlation}

\author{\IEEEauthorblockN{Simin Xu$^\ast$, Nan Yang$^\ast$, Biao~He$^\dag$, and Hamid~Jafarkhani$^\ddag$}
\IEEEauthorblockA{$^\ast$Research School of Electrical, Energy and Materials Engineering,\\Australian National University, Canberra, ACT 2601, Australia\\
$^\dag$MediaTek USA Inc., Irvine, CA 92606, USA\\
$^\ddag$Center for Pervasive Communications and
Computing, University of California, Irvine, CA 92697-2625, USA}
\IEEEauthorblockA{Email: simin.xu@anu.edu.au, nan.yang@anu.edu.au, biao.he.1990@gmail.com, hamidj@uci.edu}}

\markboth{Submitted to IEEE WCNC 2020}{Xu \MakeLowercase{\textit{et
al.}}: Coverage Analysis of Relay Assisted Millimeter Wave Cellular Networks with Spatial Correlation}

\maketitle

\begin{abstract}
We propose a novel analytical framework for evaluating the coverage performance of a millimeter wave (mmWave) cellular network where idle user equipments (UEs) act as relays. In this network, the base station (BS) adopts either the direct mode to transmit to the destination UE, or the relay mode if the direct mode fails, where the BS transmits to the relay UE and then the relay UE transmits to the destination UE. To address the drastic rotational movements of destination UEs in practice, we propose to adopt selection combining at destination UEs. New expression is derived for the signal-to-interference-plus-noise ratio (SINR) coverage probability of the network. Using numerical results, we first demonstrate the accuracy of our new expression. Then we show that ignoring spatial correlation, which has been commonly adopted in the literature, leads to severe overestimation of the SINR coverage probability. Furthermore, we show that introducing relays into a mmWave cellular network vastly improves the coverage performance. In addition, we show that the optimal BS density maximizing the SINR coverage probability can be determined by using our analysis.
\end{abstract}

\section{Introduction}\label{sec:introduction}

Millimeter wave (mmWave) has been widely acknowledged as a key enabler for multi-gigabit-per-second cellular networks \cite{overview3}, since the rich available spectrum at the mmWave band can effectively resolve the bandwidth shortage problem faced by global cellular operators. One of the most pressing challenges for mmWave cellular networks is to enhance the coverage performance, due to the unique properties of mmWave propagation such as its extreme sensitivity to blockage \cite{overview4}. One promising strategy to enhance the coverage performance is to introduce relays into mmWave cellular networks for circumventing the blockage \cite{relay3,relay1}. The introduction of relays can be realized by either deploying infrastructure relays or allowing idle user equipments (UEs) to function as relays (e.g., idle UEs serve as device-to-device (D2D) transmitters) \cite{relay1}. In this paper, we consider idle relay UEs since such relays have practical advantages relative to infrastructure relays, e.g., avoiding extra cost in infrastructure.


Motivated by the benefits of relays, the performance of relay assisted mmWave networks has been studied in the literature, e.g., \cite{relay5,relay4,relay6,relay7,relay1}. For example, \cite{relay5} investigated relay selection among multiple decode-and-forward (DF) relays between one base station (BS) and one UE. Focusing on an amplified-and-forward (AF) relay, \cite{relay4} evaluated the maximum achievable rates for half-duplex and full-duplex relaying. It is noted that neither \cite{relay5} nor \cite{relay4} considered multiple BSs or multiple UEs. To overcome such limitations, \cite{relay6} investigated the coverage performance of a relay assisted mmWave cellular network with multiple BSs and multiple UEs while ignoring the interference in the network. Recently, \cite{relay1} considered the interference when analyzing the coverage performance of a relay assisted mmWave cellular network. However, small scale fading was not considered in this analysis and directional combining was assumed at UEs.

The consideration of directional combining at multi-antenna UEs, such as \cite{relay1}, was commonly adopted in mmWave studies, e.g., \cite{cover3,BSCoop,cover6,SiminICC18}. The main rationale behind this consideration is that directional combining not only enhances the desired signal quality but also reduces the interference, under the assumption that the direction of the UE is perfectly aligned with its associated transmitter. However, as pointed out in \cite{Globecom181}, it is extremely difficult to keep this perfect alignment in practical usage scenarios, due to the frequent drastic rotational movements of UEs, especially when UEs are not idle. For example, when people play video games, the UE is frequently rotated since more and more video games need to exploit the gyroscopic sensor at the UE. Notably, the rotational movement when people play video games can be up to 80$^{\circ}$ per 100 ms \cite{Globecom181}. When the UE adopts directional combining, its rotational movement causes a severe misalignment problem, which significantly damages the reliability of the mmWave cellular network \cite{Globecom181}. Motivated by this, we propose to adopt \emph{selection combining} at UEs when they receive downlink data. We emphasize that, unlike directional combining, selection combining does not incur misalignment problems since it does not require the UE to align its direction with its associated transmitter.


In this paper, we analyze the coverage performance of a relay assisted mmWave cellular network which consists of multiple BSs, multiple relay UEs, and multiple destination UEs. In this network, the BS transmits data either directly to the destination UE, or indirectly via the relay UE when the direct transmission fails.  We propose a new analytical framework to derive the signal-to-interference-plus-noise ratio (SINR) coverage probability of the network.  In this framework, we consider (1) generalized Nakagami-$m$ fading, (2) interference, (3) selection combining at destination UEs, and (4) \emph{spatial correlation}. Here, spatial correlation addresses the phenomenon that the received signals at the destination UE antennas are not independent, which has commonly been ignored in existing studies for simplifying analysis. Using numerical results, we first demonstrate the correctness of the derived SINR coverage probability. We also show that ignoring spatial correlation leads to severe overestimation of the coverage probability, as well as severe underestimation of the minimum number of antennas needed at destination UEs to achieve an SINR coverage probability target. We further show the benefit of introducing relays for coverage performance enhancement and that the optimal BS density maximizing the SINR coverage probability can be found through our analysis.

\section{Relay Assisted MmWave Cellular Network}\label{sec:system_model}

\begin{figure}[!t]
    \begin{center}
       \includegraphics[height=2.5in]{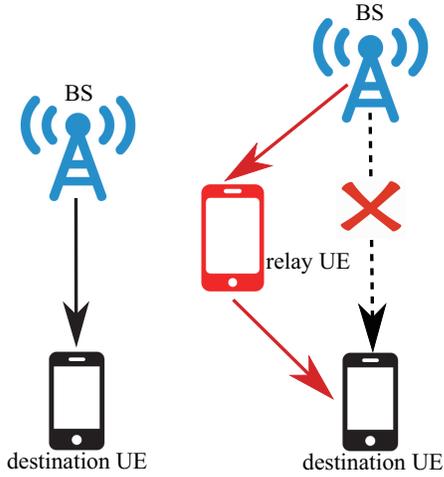}
        \caption{The two transmission modes of the considered relay assisted mmWave cellular network.}
        \label{fig:SystemModel}\vspace{-2em}
    \end{center}
\end{figure}

In this paper, we consider a relay assisted mmWave cellular network where BSs communicate with UEs. In this network, we assume that there are three types of UEs: 1) Destination UEs which are the UEs that require downlink data from BSs, 2) Uplink UEs which are the UEs that transmit uplink data to BSs, and 3) Relay UEs which are idle UEs that function as relays (or equivalently, D2D transmitters) to help the transmission between a BS and a destination UE. As depicted in Fig.~\ref{fig:SystemModel}, there are two transmission modes in the considered network: 1) Direct mode where the BS transmits data to the destination UE directly, and 2) Relay mode where the BS first transmits data to the relay UE and then the relay UE forwards data to the destination UE. Specifically, the relay mode serves as a backup option when the direct mode fails, i.e., the SINR of the direct link from the BS to the destination UE is below a given SINR threshold.

\subsection{Locations of BSs and UEs}

We assume that the locations of BSs, relay UEs, and destination UEs follow three PPPs in $\mathbb{R}^2$, the densities of which are denoted by $\hat{\lambda}_{b}$, $\hat{\lambda}_{r}$, and $\hat{\lambda}_{d}$, respectively.
In this paper, we randomly select one destination UE, refer to it as the typical destination UE, and establish a polar coordinate system with the typical destination UE at the origin, $\textbf{o}$. Based on the Slivnyak theorem, the conclusions drawn for the typical destination UE can be extended to other destination UEs.

We note that mmWave communications are extremely sensitive to blockage \cite{overview4}. Moreover, the diffraction of mmWave propagation is weak due to the used high frequencies. Thus, in this work we only consider line-of-sight (LoS) links as in \cite{cover6,cover7}. To model the spatial distribution of LoS BSs and LoS relay UEs, we adopt an accurate and simple LoS ball model as in \cite{SiminICC18,cover6,hetnet7}. As per this model, the probability of a link being LoS, $P_{\textrm{LoS}}\left(\ell\right)$, is a function of the distance between the transmitter and the receiver, $\ell$. Mathematically, $P_{\textrm{LoS}}\left(\ell\right)$ is given by
\begin{equation}\label{LoS_ball_model}
P_{\textrm{LoS}}\left(\ell\right)=
\begin{cases}
\varrho_{\varepsilon}, & \mbox{if $0<\ell<r_{\varepsilon}$}, \\
0, & \mbox{otherwise},
\end{cases}
\end{equation}
where $\varepsilon\in \left\{b,u\right\}$, $r_{b}$ and $r_{u}$ are the radii of the BS LoS ball and the relay UE LoS ball, respectively, and $0\leq\varrho_{b}\leq1$ and $0\leq\varrho_{u}\leq1$ are the LoS probabilities of BSs and relay UEs, respectively. According to the thinning theorem \cite[Prop. (1.3.5)]{Book1}, the LoS BSs follow a PPP $\Phi_{b}$ with density $\lambda_{b}\triangleq\varrho_{b}\hat{\lambda}_{b}$ within the circular area $\mathcal{B}\left(\textbf{o},r_{b}\right)$. Similarly, the LoS relay UEs follow another PPP $\Phi_{r}$ with density $\lambda_{r}\triangleq\varrho_{u}\hat{\lambda}_{r}$ within the circular area $\mathcal{B}\left(\textbf{o},r_{u}\right)$. We denote the set of LoS BSs by $\Phi_{b}=\left\{b_0,b_1,b_2,\ldots\right\}$, where $b_0$ is the associated BS, $b_k$ is the $k$th interfering BS, and $k\in\{1,2,\ldots\}$.

\subsection{Directional Gains}

We assume that an $N_{b}$ element uniform linear antenna array is adopted at each BS and an $N_{u}$ element uniform linear antenna array is adopted at each UE. At BSs, highly directional beams are assumed to be adopted to combat the high path loss of mmWave transmission. Then we note that destination UEs often experience rapid rotational movements. Thus, we assume that the typical destination UE does not leverage directional beams to avoid the beam misalignment caused by rapid rotational movements. Instead, it adopts selection combining to choose the signal with the maximum SINR, out of $N_u$ received SINRs at $N_u$ antennas. Furthermore, we assume that relay UEs leverage directional beams since relay UEs are idle such that they do not experience rapid rotational movements.

To characterize the mmWave transmission in the network, we adopt the sectored directional gain model \cite{hetnet9,hetnet7,blockModel2} which incorporates key characteristics including the beamwidth, main lobe gain, and side lobe gain. Under the assumption that an $N_{b}$ element uniform linear antenna array is adopted at each BS, the beamwidth, main lobe gain, and side lobe gain of BSs are given by $\theta_b=\frac{102 \pi}{180 N_{b}}$, $G_{B}=N_{b}$, and $G_{b}=\frac{1}{N_{b}}$, respectively. Similarly, the beamwidth, main lobe gain, and side lobe gain of relay UEs and uplink UEs are given by $\theta_u=\frac{102 \pi}{180 N_{u}}$, $G_{U}=N_{u}$, and $G_{u}=\frac{1}{N_{u}}$, respectively. Based on such notations, the directional gain of $b_{k}$, $G_{b_{k}}$, is a function of the angle off the boresight direction, $\theta_o$, given by $G_{b_{k}}=G_{B}$, if $|\theta_o|\leq\frac{1}{2}\theta_{b}$, and $G_{b_{k}}=G_{b}$, otherwise. Here, we model $\theta_o$ as a uniform random variable between $-\pi$ and $\pi$. Thus, $G_{b_{k}}$ is a discrete random variable which equals $G_{B}$ with probability (w. p.) $\frac{\theta_{b}}{2\pi}$, and $G_{b}$ w. p. $1-\frac{\theta_{b}}{2\pi}$. Similarly, the directional gain of relay UEs, $G_{e}$, is a discrete random variable which equals $G_{U}$ w. p. $\frac{\theta_{u}}{2\pi}$, and $G_{u}$ w. p. $1-\frac{\theta_{u}}{2\pi}$. Furthermore, we define the directional gain from $b_{k}$ to the relay UE as $G_{b_{k}r} \triangleq G_{b_{k}}G_{e}$. Thus, the probability mass function of $G_{b_{k}r}$ is given by
\begin{equation}\label{eq:brGain}
G_{b_{k}r}=
\begin{cases}
G_{B}G_{U}, & \mbox{w. p. $\frac{\theta_{b}\theta_{u}}{4\pi^2}$}, \\
G_{B}G_{u}, & \mbox{w. p. $\frac{\theta_{b}}{2\pi}\left(1-\frac{\theta_{u}}{2\pi}\right)$},\\
G_{b}G_{U}, & \mbox{w. p. $\left(1-\frac{\theta_{b}}{2\pi}\right)\frac{\theta_{u}}{2\pi}$}, \\
G_{b}G_{u}, & \mbox{w. p. $\left(1-\frac{\theta_{b}}{2\pi}\right)\left(1-\frac{\theta_{u}}{2\pi}\right)$}.
\end{cases}
\end{equation}

\subsection{Association Strategy}

We assume that all BSs have the same transmit power, $P_{b}$, and all relay UEs and uplink UEs have the same transmit power, $P_{u}$. We then assume that the maximum signal strength association strategy is applied in the network. Thus, in the direct mode, the destination UE always associates with its nearest LoS BS, while in the relay mode, the destination UE always associates with its nearest LoS relay UE and the associated relay UE always associates with the associated relay UE's nearest LoS BS \cite{relay1}. Thus, throughout this paper, the link from the associated BS to the typical destination UE is referred to as the \emph{direct link}, the link from the associated BS to the associated relay UE is referred to as the \emph{BR link}, and the link from the associated relay UE to the typical destination UE is referred to as the \emph{RD link}. We denote the distances of the direct link, the BR link, and the RD link by $\ell_{bd}$, $\ell_{br}$, and $\ell_{rd}$, respectively. If $\Phi_{b}\neq\emptyset$, the probability density function (PDF) of $\ell_{bd}$ is the same as the PDF of $\ell_{br}$, which is given by $f_{\ell_{bd}}\left(x\right)=f_{\ell_{br}}\left(x\right)=2\pi\lambda_{b}{}xe^{-\pi\lambda_{b} x^2}/{\Xi_{b}}$, where $\Xi_{b}\triangleq \Pr\left(\Phi_{b}\neq \emptyset\right)=1-e^{-\pi\lambda_{b}{}r_{b}^{2}}$ and $0<x<r_{b}$ \cite{Book1}. If $\Phi_{r}\neq\emptyset$, the PDF of $\ell_{rd}$ is given by $f_{\ell_{rd}}\left(x\right)=2\pi\lambda_{r}{}xe^{-\pi\lambda_{r} x^2}/{\Xi_{r}}$, where $\Xi_{r}\triangleq \Pr\left(\Phi_{r}\neq \emptyset\right)=1-e^{-\pi\lambda_{r}{}r_{u}^{2}}$ and $0<x<r_{u}$.

\subsection{SINR Formulation}

In this subsection, we first characterize the interference from the BS to the destination UE, the interference from the BS to the relay UE, and the interference from the relay UE to the destination UE. Then we formulate the SINRs between two nodes based on the interference characterization.

We assume that frequency division duplex is adopted, i.e., downlink frequency band and uplink frequency band are non-overlapping. Thus, the transmission using the downlink frequency band and the transmission using the uplink frequency band do not interfere with each other. In this network, we clarify that there are three types of transmitters: 1) BSs that transmit to destination UEs in the direct mode and relay UEs in the relay mode, 2) Relay UEs that transmit to destination UEs, and 3) Uplink UEs that transmit uplink data to BSs. Similar to \cite{relay1}, we assume that BSs transmit using the downlink frequency band, while relay UEs and uplink UEs transmit using the uplink frequency band. Thus, the transmission from BSs and the transmissions from relay UEs and uplink UEs do not interfere with each other, but the transmission from relay UEs and the transmission from uplink UEs interfere with each other. It follows that in the direct link and the BR link, the interference solely comes from other BSs. Differently, in the RD link, the interference comes from not only other relay UEs but also uplink UEs.

We emphasize that not all relay UEs and uplink UEs interfere with the RD link. 
It is widely acknowledged that in a cellular network, network resources must be divided into multiple sub-channels to enable multiple access.
In this paper, we assume that 
the UEs using the same sub-channel interfere with each other, while the UEs using different sub-channels do not. Thus, in the RD link, only the relay UEs and uplink UEs which use the same sub-channel as the associated relay UE cause interference. In the following, we define all interfering relay UEs and uplink UEs as interfering UEs. In order to model the distribution of all interfering UEs, we define a multiplexing factor, $\rho$, which represents the average number of interfering UEs on each sub-channel within the cell \cite{relay1}. Considering the distribution of BSs, the distribution of LoS interfering UEs can be modeled by a homogeneous PPP with density $\lambda_{i}\triangleq \varrho_{u} \rho \hat{\lambda}_{b}$ within the circular area $\mathcal{B}\left(\textbf{o},r_{u}\right)$, $\Phi_{i}=\left\{i_1,i_2,\ldots\right\}$, where $i_{k}$ is the $k$th interfering UE and $k\in\{1,2,\ldots\}$.



Based on the aforementioned interference characterization, we obtain that the SINR at antenna $n$ in the direct link, denoted by $\gamma_{bd,n}$ where $n\in\left\{1,2,\ldots,N_u\right\}$, is written as
\begin{equation}\label{eq:SINRbd}
\gamma_{bd,n}=\frac{P_{b}G_{B}\ell_{bd}^{-\eta}\left|h_{nb_0}\right|^{2}}
{\sum_{b_k\in\tilde{\Phi}_{b}}P_{b}G_{b_k}\ell_{b_{k}d}^{-\eta}\left|h_{nb_k}\right|^{2}+\sigma^{2}},
\end{equation}
where
$\tilde{\Phi}_{b}\triangleq\Phi_{b}\backslash\left\{b_0\right\}$ denotes the set of LoS BSs excluding $b_{0}$, $\ell_{b_{k}d}$ is the distance between $b_{k}$ and the typical destination UE, $\eta \ge 2$ is the path loss exponent, $h_{nb_{l}}$ is the small scale fading from $b_{l}$ to antenna $n$, $l\in\{0,k\}$, and $\sigma^{2}$ is the variance of the additive white Gaussian noise (AWGN) at antenna $n$.
In this work, we assume that $h_{nb_{l}}$ is subject to independent and identically distributed (i.i.d.) Nakagami-$m$ fading\footnote{We clarify that Nakagami-$m$ fading is a generalized fading model which can fit a variety of empirical measurements, such as Rayleigh fading by setting $m=1$ and Rician-$K$ fading by setting $m=\left(K+1\right)^2/\left(2K+1\right)$.}. By defining $\hbar_{nb_{l}} \triangleq\left|h_{nb_{l}}\right|^{2}$, we find that $\hbar_{nb_{l}}$ follows the Gamma distribution with the shape parameter $m_{bd}$ and the scale parameter $\frac{1}{m_{bd}}$, i.e., $\hbar_{nb_{l}} \sim\Gamma\left(m_{bd},\frac{1}{m_{bd}}\right)$.

In the BR link, the SINR at the relay UE, denoted by $\gamma_{br}$, is written as
\begin{equation}\label{eq:SINRbr}
\gamma_{br}=\frac{P_{b}G_{B}G_{U}\ell_{br}^{-\eta} \hbar_{b_0}}
{\sum_{b_k\in\tilde{\Phi}_{b}}P_{b}G_{b_{k}r}\ell_{b_{k}r}^{-\eta} \hbar_{b_k}+\sigma^{2}},
\end{equation}
where $\ell_{b_{k}r}$ is the distance between $b_{k}$ and the relay UE, $\hbar_{b_{l}}$ is the small scale fading gain between $b_{l}$ and the relay UE, $l\in\{0,k\}$. In this work, we assume that $\hbar_{b_{l}}$ is subject to i.i.d. Gamma distribution with the shape parameter $m_{br}$ and the scale parameter $\frac{1}{m_{br}}$, i.e., $\hbar_{b_{l}} \sim\Gamma\left(m_{br},\frac{1}{m_{br}}\right)$.

In the RD link, the SINR at antenna $n$, $n\in\left\{1,2,\ldots,N_u\right\}$, denoted by $\gamma_{rd,n}$, is written as
\begin{equation}\label{eq:SINRrd}
\gamma_{rd,n}=\frac{P_{u}G_{U}\ell_{rd}^{-\eta}\hbar_{nr} }
{\sum_{i_{k}\in\Phi_{i}}P_{u}G_{e}\ell_{i_{k}d}^{-\eta}\hbar_{ni_k} +\sigma^{2}},
\end{equation}
where $\ell_{i_{k}d}$ is the distance between the $k$th interfering UE and the destination UE, $\hbar_{nr}$ is the small scale fading gain from the relay UE to antenna $n$, $\hbar_{ni_k}$ is the small scale fading from $i_{k}$ to antenna $n$.
In this work, we assume that $\hbar_{nr}$ and $\hbar_{ni_k}$ are subject to i.i.d. Gamma distribution with the shape parameter $m_{rd}$ and the scale parameter $\frac{1}{m_{rd}}$, i.e., $\hbar_{nr}\sim\Gamma\left(m_{rd},\frac{1}{m_{rd}}\right)$ and $\hbar_{ni_k}\sim\Gamma\left(m_{rd},\frac{1}{m_{rd}}\right)$.

Under the assumption that the Nakagami-$m$ fading parameters, e.g., $m_{bd}$, $m_{br}$, and $m_{rd}$, are positive integers \cite{hetnet9,cover6}, the moment generating function (MGF) of $\hbar_{\mu}$ is given by $M_{\hbar_{\mu}}\left(s\right)=\left(1-\frac{s}{m_{\nu}}\right)^{-m_{\nu}}$, where $(\mu,\nu)\in\{(nb_{l},~m_{bd}),(b_{l},~m_{br}),(nr,~m_{rd}),(ni_{k},~m_{rd})\}$. Moreover, as given in \cite[Lemma 1]{GammaCDFLB}, the cumulative distribution function (CDF) of $\hbar_{\mu}$ can be tightly lower bounded by
\begin{equation}\label{eq:lower_bound_Gamma}
F_{\hbar_{\mu}}\left(x\right)
=1-e^{-m_{\nu}x}\sum_{\omega=0}^{m_{\nu}-1}\frac{\left(m_{\nu}x\right)^{\omega}}{\omega!}
>\left(1-e^{-\alpha_{\nu}{}x}\right)^{m_{\nu}},
\end{equation}
where $\alpha_{\nu}=m_{\nu}\left(m_{\nu}!\right)^{-\frac{1}{m_{\nu}}}$. In the analysis presented in Section \ref{sec:coverage_analysis}, we adopt the lower bound on $F_{\hbar_{\mu}}\left(x\right)$ since this lower bound enables us to derive the SINR coverage probability when the Nakagami-$m$ fading parameters are higher than one.


By observing \eqref{eq:SINRbd} and \eqref{eq:SINRrd}, we find that the interference signals at the $N_u$ UE antennas are correlated due to the common interfering BSs' location or the common interfering UEs' location. Also, the desired signals at the $N_u$ UE antennas are correlated due to the common associated BS' location or the common associated relay UE's location. Despite that both interference signals and desired signals are correlated in wireless networks, these spatial correlations are often ignored in the existing studies on mmWave systems to simplify the analysis. To overcome this, we address the spatial correlation when analyzing the coverage performance of the network considered in this paper.

\section{SINR Coverage Probability Analysis}\label{sec:coverage_analysis}

In this section, we derive the SINR coverage probability of the considered network, $\mathbb{P}_{N_u}\left(\tau\right)$. Here, $\mathbb{P}_{N_u}\left(\tau\right)$ is defined as the probability that the SINR at the $N_u$-antenna typical destination UE is higher than a given SINR threshold, $\tau$. As aforementioned, we derive $\mathbb{P}_{N_u}\left(\tau\right)$ by considering spatial correlation at the destination UE. This mandates a new and challenging analytical framework for the SINR coverage probability, as pointed out in \cite{Globecom182}. Furthermore, we aim to answer an open question in mmWave cellular networks: ``\emph{Is the analysis error caused by ignoring spatial correlation acceptable?}'' To answer this question, we also derive the SINR coverage probability for the case where spatial correlation is ignored, which enables us to compare the two coverage probabilities in Section~\ref{sec:numerical_results}.

We assume that the half-duplex DF relay protocol is adopted. Thus, in the relay mode, the SINR at the typical destination UE is higher than $\tau$ when both the SINR of the BR link and the SINR of the RD link are higher than $\tau$. We denote the SINR coverage probability of the direct link, the BR link, and the RD link as $\mathbb{P}_{N_u,bd}\left(\tau\right)$, $\mathbb{P}_{br}\left(\tau\right)$, and $\mathbb{P}_{N_u,rd}\left(\tau\right)$, respectively. This allows us to express $\mathbb{P}_{N_u}\left(\tau\right)$ as
\begin{equation}\label{eq:coverage_probability}
\mathbb{P}_{N_u}\left(\tau\right)=1-\left[1-\mathbb{P}_{N_u,bd}\left(\tau\right)\right]
\left[1-\mathbb{P}_{br}\left(\tau\right)\mathbb{P}_{N_u,rd}\left(\tau\right)\right].
\end{equation}
We next derive and present $\mathbb{P}_{N_u,bd}\left(\tau\right)$, $\mathbb{P}_{br}\left(\tau\right)$, and $\mathbb{P}_{N_u,rd}\left(\tau\right)$ in Theorems \ref{theorem:spatial_correlation_bd}, \ref{theorem:br}, and \ref{theorem:spatial_correlation_rd}, respectively.

\newtheorem{theorem}{Theorem}
\begin{theorem}\label{theorem:spatial_correlation_bd}
Under the assumption that 
$F_{\hbar_{nb_{l}}}\left(x\right)$ is approximated as $\left(1-e^{-\alpha_{bd}{}x}\right)^{m_{bd}}$, the SINR coverage probability of the direct link is derived as
\begin{align}\label{eq:coverage_probability_bd}
&\mathbb{P}_{N_u,bd}\left(\tau\right)
=\Xi_b\sum\limits_{\kappa=1}^{N_u}\left(-1\right)^{\kappa+1}\binom{N_u}{\kappa}  \sum\limits_{j_1+j_2+\cdots+j_{m_{bd}}=\kappa}\beta_{bd}\notag\\
&\times{}\int_{0}^{r_b}e^{-\frac{\tau\alpha_{bd}\Omega_{bd}\sigma^{2}}{P_b G_{B}x^{-\eta}}-2\pi\lambda_b\int_{x}^{r_b} \ell \left(1-v_{bd}\left(\ell \right)\right)d\ell }f_{\ell_{bd}}\left(x\right)dx,
\end{align}
where $j_1,j_2, \ldots, j_{m_{bd}}$ are nonnegative integers, $\Omega_{bd}\triangleq j_1+2j_2+\cdots+m_{bd} j_{m_{bd}}$, $\delta_{bd}\triangleq\frac{\alpha_{bd}\tau}{P_bG_{B}\ell_{bd}^{-\eta}}$, $\beta_{bd}$ is defined as
\begin{align}
\beta_{bd}&\triangleq\binom{\kappa}{j_1}\binom{\kappa-j_1}{j_2} \cdots
\binom{\kappa-j_1-\cdots-j_{m_{bd}-1}}{j_{m_{bd}}}
\binom{m_{bd}}{1}^{j_1}\notag\\
&\hspace{-4mm}\times \binom{m_{bd}}{2}^{j_2}\cdots\binom{m_{bd}}{m_{bd}}^{j_{m_{bd}}}
\left(-1\right)^{j_1+j_2+\cdots+j_{m_{bd}}+\Omega_{bd}},
\end{align}
and $v_{bd}\left(\ell\right)$ is defined as
\begin{align}\label{eq:v_definition}
v_{bd}\left(\ell \right)&\triangleq\hspace{0mm}\frac{\theta_{b}}{2\pi}\prod_{q=1}^{m_{bd}}
\left(1+\frac{q\delta_{bd}{}P_b G_{B}\ell^{-\eta}}{m_{bd}}\right)^{-m_{bd}j_{q}}\notag\\
&\hspace{0mm}+\left(1-\frac{\theta_{b}}{2\pi}\right)\prod_{q=1}^{m_{bd}}
\left(1+\frac{q\delta_{bd}{}P_b G_{b}\ell^{-\eta}}{m_{bd}}\right)^{-m_{bd}j_{q}}.
\end{align}
\begin{IEEEproof}
We define $\epsilon_n$ as the event that $\gamma_{bd,n}$ is higher than $\tau$, where $n\in\left\{1,2,\ldots,N_u\right\}$. As per the rules of selection combining, $\mathbb{P}_{N_u,bd}\left(\tau\right)$ is written as
\begin{align}\label{eq:intersection}
&\mathbb{P}_{N_u,bd}\left(\tau\right)=\Pr\left(\bigcup\limits_{n=1}^{N_u}\epsilon_n\right)
\notag\\&=\sum\limits_{\kappa=1}^{N_u}\left(-1\right)^{\kappa+1}\hspace{-5.5mm}\sum_{1\leq\zeta_1<\zeta_2< \cdots<\zeta_\kappa\leq N_u}\hspace{-5.5mm}\Pr\left(\epsilon_{\zeta_1}\cap\epsilon_{\zeta_2}\cap\cdots\cap \epsilon_{\zeta_\kappa}\right).
\end{align}

We note that in the considered network, the probability of the intersection of $\kappa$ events, $\Pr\left(\epsilon_{\zeta_1} \cap \epsilon_{\zeta_2}\cap \cdots \cap \epsilon_{\zeta_\kappa}\right)$, is the same, regardless of which $\kappa$ out of $N_u$ events are chosen. Thus, we rewrite \eqref{eq:intersection} as
\begin{equation}\label{eq:union_to_intersection}
\mathbb{P}_{N_u,bd}\left(\tau\right)=\sum\limits_{\kappa=1}^{N_u} \left(-1\right)^{\kappa+1} \binom{N_u}{\kappa}\Pr\left(\bigcap\limits_{n=1}^{\kappa}\epsilon_n\right).
\end{equation}

Based on 
\eqref{eq:SINRbd}, we derive $\Pr\left(\bigcap\limits_{n=1}^{\kappa}\epsilon_n\right)$ as
\begin{align}\label{eq:union_calculation}
&\hspace{-4mm}\Pr\left(\bigcap\limits_{n=1}^{\kappa}\epsilon_n\right)\notag\\
=&\Xi_b\Pr\left(\bigcap\limits_{n=1}^{\kappa}\hbar_{nb_0}>\frac{\tau \left(\sum_{b_k\in\tilde{\Phi}}P_b G_{b_k}\ell_{b_{k}d}^{-\eta}\hbar_{nb_k}+\sigma^{2}\right)}
{P_b G_{B}\ell_{bd}^{-\eta}}\right)\notag\\
\stackrel{\left(f\right)}{=}&
\Xi_b \mathbb{E}_{\ell_{bd},\tilde{\Phi}_{b},G_{b_k},\vartheta}\Bigg[\prod_{n=1}^{\kappa} \sum\limits_{j=1}^{m_{bd}}\binom{m_{bd}}{j}\left(-1\right)^{j+1}\notag\\
&\hspace{24mm}\times{}e^{-j\delta_{bd}\left(\sum_{b_k\in\tilde{\Phi}}P_b G_{b_k}\ell_{b_{k}d}^{-\eta}\hbar_{nb_k}+\sigma^{2}\right)}\Bigg].
\end{align}
Here, step $\left(f\right)$ is achieved by following the binomial theorem and making the assumption that $F_{\hbar_{nb_{l}}}\left(x\right)$ is approximated as $\left(1-e^{-\alpha_{bd}{}x}\right)^{m_{bd}}$, while $\vartheta$ is defined as $\vartheta\triangleq\sum_{q=1}^{m_{bd}}\sum_{p=1}^{j_{q}}q \hbar_{I_{q,p},b_k}$,
where $I_{q,p}\in\{1,\ldots,\kappa\}$ with $q\in\{1,\ldots,m_{bd}\}$ and $p\in\{1,\ldots,j_{q}\}$ and the values of $I_{q,p}$ differ from each other.
By defining 
$\Psi=\sum_{b_k\in\tilde{\Phi}_{b}}\delta_{bd}{}P_b G_{b_k}\ell_{b_{k}d}^{-\eta}\vartheta$, we further derive \eqref{eq:union_calculation} as
\begin{align}\label{eq:SINR_derivation}
&\Pr\left(\bigcap\limits_{n=1}^{\kappa}\epsilon_n\right)\notag\\
&=\Xi_b\hspace{-5.5mm} \sum_{j_1+j_2+\cdots+j_{m_{bd}}=\kappa}\hspace{-5.5mm}\mathbb{E}_{\ell_{bd}}\bigg[e^{-\delta_{bd}\sigma^{2}\Omega_{bd}}
\sum_{I_{q,p}}\mathbb{E}_{\tilde{\Phi}_{b},G_{b_k},\vartheta}\Big[e^{-\Psi}\Big]\bigg]
\notag\\
&\times\binom{m_{bd}}{1}^{j_1}\hspace{-0.5mm}\binom{m_{bd}}{2}^{j_2}\hspace{-1mm}
\cdots\binom{m_{bd}}{m_{bd}}^{j_{m_{bd}}}
\hspace{-2mm}\left(-1\right)^{j_1+j_2+\cdots+j_{m_{bd}}+\Omega_{bd}}
\notag\\
&\stackrel{\left(g\right)}{=}\Xi_b\hspace{-5.5mm}\sum_{j_1+j_2+\cdots+j_{m_{bd}}=\kappa}\hspace{-5.5mm}
\beta_{bd}\mathbb{E}_{\ell_{bd}}\bigg[e^{-\delta_{bd}\sigma^{2}\Omega_{bd}}
\mathbb{E}_{\tilde{\Phi}_{b},G_{b_k},\vartheta}\Big[e^{-\Psi}\Big]\bigg].
\end{align}
To achieve step $\left(g\right)$, we find that given 
$M_{\hbar_{nb_l}}\left(s\right)$, the MGF of $\vartheta$ is $M_{\vartheta}\left(s\right)=\prod_{q=1}^{m_{bd}}\left(1-\frac{qs}{m_{bd}}\right)^{-m_{bd}j_{q}}$. This indicates that given $j_1,j_2,\ldots,j_{m_{bd}}$, the MGF of $\vartheta$ does not change with the values of $I_{q,p}$.
We also find that given $j_1,j_2,\ldots,j_{m_{bd}}$, $\mathbb{E}_{\tilde{\Phi},G_{b_k},\vartheta}\big[e^{-\Psi}\big]$ does not change with the values of $I_{q,p}$.
Thus, step $\left(g\right)$ is achieved due to $\sum_{I_{q,p}}\mathbb{E}_{\tilde{\Phi}_b,G_{b_k},\vartheta}\big[e^{-\Psi}\big]
=\binom{k}{j_1}\binom{k-j_1}{j_2}\cdots\binom{k-j_1-\cdots-j_{m_{bd}-1}}{j_{m_{bd}}} \mathbb{E}_{\tilde{\Phi}_b,G_{b_k},\vartheta}\big[e^{-\Psi}\big]$.

Using the MGF of $\vartheta$, we derive $\mathbb{E}_{\tilde{\Phi}_b,G_{b_k},\vartheta}\big[e^{-\Psi}\big]$ as
\begin{align}\label{eq:expectation}
&\mathbb{E}_{\tilde{\Phi}_b,G_{b_k},\vartheta}\big[e^{-\Psi}\big]
\notag\\&=\mathbb{E}_{\tilde{\Phi}_b,G_{b},\vartheta}
\Bigg[\prod_{\tilde{\Phi}_b}e^{-\delta_{bd}{}P_bG_{b_{k}}\ell_{b_{k}d}^{-\eta}\vartheta}\Bigg]\notag\\
&\stackrel{\left(h\right)}{=}\mathbb{E}_{\tilde{\Phi}_b,G_{b_k}}
\Bigg[\prod_{\tilde{\Phi}_b}\prod_{q=1}^{m_{bd}}
\left(1+\frac{q\delta_{bd}{}P_b G_{b_{k}}\ell_{b_{k}d}^{-\eta}}{m_{bd}}\right)^{-m_{bd}j_{q}}\Bigg]\notag\\
&=\mathbb{E}_{\tilde{\Phi}_b}\Bigg[\prod_{\tilde{\Phi}_b}v_{bd}\left(\ell_{b_{k}d}\right)\Bigg]
\notag\\&=e^{-2\pi\lambda_b\int_{\ell_{bd}}^{r_b}\ell \left(1-v_{bd}\left(\ell \right)\right)d\ell}.
\end{align}
Here, step $\left(h\right)$ is achieved by using $M_{\vartheta}\left(s\right)$. Finally, by substituting \eqref{eq:SINR_derivation} and \eqref{eq:expectation} into \eqref{eq:union_to_intersection}, we obtain the desired result in \eqref{eq:coverage_probability_bd} and complete the proof.
\end{IEEEproof}
\end{theorem}

\begin{theorem}\label{theorem:br}
Under the assumption that the CDF of $\hbar_{b_{l}}$ is approximated as $\left(1-e^{-\alpha_{br}{}x}\right)^{m_{br}}$,
the SINR coverage probability 
of the BR link is derived as
\begin{align}\label{br}
\mathbb{P}_{br}\left(\tau\right)=&\Xi_b\sum_{j=1}^{m_{br}}\left(-1\right)^{j+1}\binom{m_{br}}{j}  \notag\\
&\times\int_{0}^{r_b}e^{-\psi x^{\eta} \sigma^{2} -2 \pi \lambda_b \int_{x}^{r_b} \ell\Omega_b d\ell}f_{\ell_{br}}(x)dx,
\end{align}
where $\psi\triangleq\frac{j\alpha_{br}\tau}{P_{b}G_{B}G_{U}}$ and $\Omega_b$ is defined as \eqref{eq:Omega_b}.
\begin{figure*}[!t]
\normalsize
\begin{align}\label{eq:Omega_b}
\Omega_b\triangleq&\frac{\theta_{u}}{2\pi}\left(1-\frac{\theta_{b}}{2\pi}\right)\left[1-\left(1+\frac{\psi x^{\eta}P_{b}G_{b}G_{U}}{m_{br}\ell^{\eta}}\right)^{-m_{br}}\right]
+\left(1-\frac{\theta_{b}}{2\pi}\right)\left(1-\frac{\theta_{u}}{2\pi}\right)\left[1-\left(1+\frac{\psi x^{\eta}P_{b}G_{b}G_{u}}{m_{br}\ell^{\eta}}\right)^{-m_{br}}\right]\notag\\
&+\frac{\theta_{b}}{2\pi}\left(1-\frac{\theta_{u}}{2\pi}\right)\left[1-\left(1+\frac{\psi x^{\eta}P_{b}G_{B}G_{u}}{m_{br}\ell^{\eta}}\right)^{-m_{br}}\right]
+\frac{\theta_{b}}{2\pi}\frac{\theta_{u}}{2\pi}\left[1-\left(1+\frac{\psi x^{\eta}P_{b}G_{B}G_{U}}{m_{br}\ell^{\eta}}\right)^{-m_{br}}\right].
\end{align}\vspace{-1em}
\hrulefill
\end{figure*}
\begin{IEEEproof}
Based on \eqref{eq:SINRbr}, we derive $\mathbb{P}_{br}\left(\tau\right)$ as
\begin{align}\label{eq:Pbr_derivation}
&\mathbb{P}_{br}\left(\tau\right)\notag\\
&=\Xi_b \mathbb{E}_{\ell_{br},I_b^{\ast}}
\left[\Pr\left(\hbar_{b_0}>\frac{\tau(I_b^{\ast}+\sigma^{2})}{P_{b}G_{B}G_{U}\ell_{br}^{-\eta}}\right)\right]\notag\\
&\stackrel{(i)}{=}\Xi_b\mathbb{E}_{\ell_{br},I_b^{\ast}}
\left[\sum_{j=1}^{m_{br}}\left(-1\right)^{j+1}\binom{m_{br}}{j}e^{-\frac{j\alpha_{br}  \tau \left( I_b^{\ast}+\sigma^{2}\right)}{P_{b}G_{B}G_{U}\ell_{br}^{-\eta}}}  \right]\notag\\
&=\Xi_b\sum_{j=1}^{m_{br}}\left(-1\right)^{j+1}\binom{m_{br}}{j}\mathbb{E}_{\ell_{br},I_b^{\ast}}
\left[e^{-\frac{\psi \left( I_b^{\ast}+\sigma^{2}\right)}{\ell_{br}^{-\eta}}}  \right]\notag\\
&=\Xi_b\sum_{j=1}^{m_{br}}\left(-1\right)^{j+1}\binom{m_{br}}{j}\mathbb{E}_{\ell_{br}}
\left[e^{-\frac{\psi \sigma^{2}}{\ell_{br}^{-\eta}}} \mathbb{E}_{I_b^{\ast}}
\left[e^{-\frac{\psi I_b^{\ast}}{\ell_{br}^{-\eta}}}  \right] \right]\notag\\
&=\Xi_b\sum_{j=1}^{m_{br}}\left(-1\right)^{j+1}\binom{m_{br}}{j}
\notag\\
&\hspace{4mm}\times\int_{0}^{r_b}e^{-\psi x^{\eta} \sigma^{2}}\mathbb{E}_{I_b^{\ast}}
\left[e^{-\frac{\psi I_b^{\ast}}{x^{-\eta}}}  \right] f_{\ell_{br}}(x) dx,
\end{align}
where $I_b^{\ast}\triangleq \sum_{b_k\in\tilde{\Phi_{b}}}P_{b}G_{b_{k}r}\ell_{b_{k}r}^{-\eta} \hbar_{b_k}$. Here,
step $(i)$ is achieved by following the binomial theorem and making the assumption that the CDF of $\hbar_{b_{l}}$ is approximated as $\left(1-e^{-\alpha_{br}{}x}\right)^{m_{br}}$. We further derive $\mathbb{E}_{I_b^{\ast}}
\left[e^{-\frac{\psi I_b^{\ast}}{x^{-\eta}}}  \right]$ as
\begin{align}\label{eq:EIbAst}
&\mathbb{E}_{I_b^{\ast}}
\left[e^{-\frac{\psi I_b^{\ast}}{x^{-\eta}}}  \right]
=\mathbb{E}_{I_b^{\ast}}
\left[e^{-\psi x^{\eta}I_b^{\ast}}  \right]\notag\\
&\stackrel{(j)}{=}e^{-\lambda_{b}\int_{x}^{r_b}\ell \int_{0}^{2\pi}
\left(1-\mathbb{E}_{\hbar_{b_k}}\left[e^{-\hbar_{b_k}{}\psi x^{\eta}{}P_{b}G_{b_{k}r}\ell^{-\eta}}\right]\right)
d\theta{}d\ell}\notag\\
&\stackrel{(k)}{=}e^{-\lambda_{b}\int_{x}^{r_b}\ell\int_{0}^{2\pi}  \left(1-\left(1+\frac{\psi x^{\eta}{}P_{b}G_{b_{k}r}\ell^{-\eta}}{m_{br}}\right)^{-m_{br}}\right)
d\theta{}d\ell}\notag\\
&\stackrel{(l)}{=}e^{-2 \pi \lambda_b\int_{x}^{r_b}\ell \Omega_{b}d\ell}.
\end{align}
Here, step $(j)$ is achieved as per \cite[Cor. (2.3.2)]{Book1}, step $(k)$ is achieved by using the MGF of Gamma distribution, and step $(l)$ is achieved as per \eqref{eq:brGain}. Finally, we substitute \eqref{eq:EIbAst} into \eqref{eq:Pbr_derivation} to obtain the desired result in \eqref{br} and thus, the proof is completed.
\end{IEEEproof}
\end{theorem}

\begin{theorem}\label{theorem:spatial_correlation_rd}
Under the assumption that the CDFs of $\hbar_{nr}$ and $\hbar_{ni_k}$ are approximated as $\left(1-e^{-\alpha_{rd}{}x}\right)^{m_{rd}}$,
the SINR coverage probability 
of the RD link is derived as
\begin{align}\label{eq:coverage_probability_rd}
&\mathbb{P}_{N_u,rd}\left(\tau\right)
=\Xi_r\sum\limits_{\kappa=1}^{N_u}\left(-1\right)^{\kappa+1}\binom{N_u}{\kappa}  \sum\limits_{j_1+j_2+\cdots+j_{m_{rd}}=\kappa}\beta_{rd}\notag\\
&\times{}\int_{0}^{r_u}e^{-\frac{\tau\alpha_{rd}\Omega_{rd}\sigma^{2}}{P_u G_{U}x^{-\eta}}-2\pi\lambda_i\int_{0}^{r_u} \ell \left(1-v_{rd}\left(\ell \right)\right)d\ell }f_{\ell_{rd}}\left(x\right)dx,
\end{align}
where $j_1, j_2, \ldots, j_{m_{rd}}$ are nonnegative integers, $\Omega_{rd}\triangleq j_1+2j_2+\cdots+m_{rd}j_{m_{rd}}$, $\delta_{rd}\triangleq\frac{\alpha_{rd}\tau}{P_uG_{U}\ell_{rd}^{-\eta}}$, $\beta_{rd}$ is defined as
\begin{align}
\beta_{rd}\triangleq&\binom{\kappa}{j_1}\binom{\kappa-j_1}{j_2} \cdots
\binom{\kappa-j_1-\cdots-j_{m_{rd}-1}}{j_{m_{rd}}}
\binom{m_{rd}}{1}^{j_1}\notag\\
&\times \binom{m_{rd}}{2}^{j_2}\cdots\binom{m_{rd}}{m_{rd}}^{j_{m_{rd}}}
\left(-1\right)^{j_1+j_2+\cdots+j_{m_{rd}}+\Omega_{rd}},
\end{align}
and $v_{rd}\left(\ell\right)$ is defined as
\begin{align}\label{eq:v_definition_alt}
v_{rd}\left(\ell \right)&\triangleq\hspace{0mm}\frac{\theta_{u}}{2\pi}\prod_{q=1}^{m_{rd}}
\left(1+\frac{q\delta_{rd}{}P_u G_{U}\ell^{-\eta}}{m_{rd}}\right)^{-m_{rd}j_{q}}\notag\\
&\hspace{-1.5mm}+\left(1-\frac{\theta_{u}}{2\pi}\right)\prod_{q=1}^{m_{rd}}
\left(1+\frac{q\delta_{rd}{}P_u G_{u}\ell^{-\eta}}{m_{rd}}\right)^{-m_{rd}j_{q}}.
\end{align}
\begin{IEEEproof}
The proof of Theorem~\ref{theorem:spatial_correlation_rd} is omitted here because \eqref{eq:coverage_probability_rd} can be obtained by following the procedure presented in the proof of Theorem~\ref{theorem:spatial_correlation_bd}.
\end{IEEEproof}
\end{theorem}

We now examine the case where spatial correlation is ignored. In this case, the SINRs received at the $N_u$ antennas of the destination UE are assumed to be independent with each other. Thus, when spatial correlation is ignored, the SINR coverage probability of the direct link, $\hat{\mathbb{P}}_{N_u,bd}\left(\tau\right)$, is given by
\begin{equation}
\hat{\mathbb{P}}_{N_u,bd}\left(\tau\right)=1-\left(1-\mathbb{P}_{1,bd}\left(\tau\right)\right)^{N_u},
\end{equation}
and the SINR coverage probability of the RD link, $\hat{\mathbb{P}}_{N_u,rd}\left(\tau\right)$, is given by
\begin{equation}
\hat{\mathbb{P}}_{N_u,rd}\left(\tau\right)=1-\left(1-\mathbb{P}_{1,rd}\left(\tau\right)\right)^{N_u}.
\end{equation}
Therefore, when spatial correlation is ignored, the probability that the SINR at the $N_u$-antenna typical destination UE is higher than $\tau$, $\hat{\mathbb{P}}_{N_u}\left(\tau\right)$, is obtained as
\begin{equation}\label{eq:neglecting_spatial_correlation}
\hat{\mathbb{P}}_{N_u}\left(\tau\right)=1-\left[1-\hat{\mathbb{P}}_{N_u,bd}\left(\tau\right)\right]
\left[1-\mathbb{P}_{br}\left(\tau\right)\hat{\mathbb{P}}_{N_u,rd}\left(\tau\right)\right].
\end{equation}
The difference between \eqref{eq:coverage_probability} and \eqref{eq:neglecting_spatial_correlation} indicates that spatial correlation imposes a profound impact on the SINR coverage probability of the $N_u$-antenna destination UE. This will be further examined in Section \ref{sec:numerical_results}.

\section{Numerical Results}\label{sec:numerical_results}

\begin{table}[t]
\centering
\caption{Simulation Parameters Used in Section \ref{sec:numerical_results}}
\label{table:Simulation_Parameters}
\begin{tabular}{|l|l|}
\hline
Parameter                         & Value                  \\ \hline
BS transmit power, $P_b$          & $35~\textrm{dBm}$      \\ \hline
UE transmit power, $P_u$          & $25~\textrm{dBm}$      \\ \hline
Noise power, $\sigma^{2}$         & $0~\textrm{dBm}$       \\ \hline
Number of antennas at each BS, $N_b$          & $10$      \\ \hline
Number of antennas at each UE, $N_u$          & $4$      \\ \hline
Path loss exponent, $\eta$        & 2.4                      \\ \hline
Nakagami-$m$ fading parameters, $m_{bd}$, $m_{br}$, $m_{rd}$ & 2                     \\ \hline
BS LoS ball radius, $r_{b}$       & 100 m                  \\ \hline
UE LoS ball radius, $r_{u}$       & 20 m                   \\ \hline
BS LoS probability, $\varrho_{b}$ & 0.9                    \\ \hline
UE LoS probability, $\varrho_{u}$ & 0.63                   \\ \hline
LoS BS density, $\lambda_{b}$     & $ 0.0002/\textrm{m}^2$ \\ \hline\
LoS relay UE density, $\lambda_{r}$  & $ 0.002/\textrm{m}^2$  \\ \hline
Multiplexing factor, $\rho$       & 0.9                    \\ \hline
\end{tabular}
\end{table}

In this section, we present numerical results to evaluate the impact of spatial correlation and network parameters on the SINR coverage probability, $\mathbb{P}_{N_u}\left(\tau\right)$. The values of the parameters used in this section are summarized in Table~\ref{table:Simulation_Parameters}, unless otherwise specified.

\begin{figure}[!t]
    \begin{center}
        \includegraphics[height=2.8in,width=0.95\columnwidth]{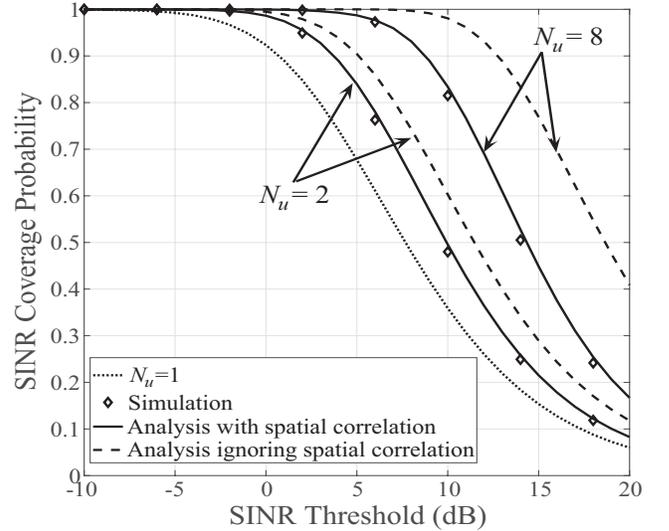}
        \caption{The SINR coverage probability, $\mathbb{P}_{N_u}\left(\tau\right)$, versus the SINR threshold, $\tau$, for different values of $N_u$.}
        \label{fig:three_lines}
    \end{center}
\end{figure}

\begin{table}[t]
\centering
\caption{$N_{\min}$ versus $\hat{N}_{\min}$ for Different Values of $\xi$}
\label{table:number_of_antennas_needed}
\begin{tabular}{|l|l|l|l|l|}
\hline
$\xi$   & 60\% & 70\% & 80\% & 90\%  \\ \hline
$N_{\min}$ & 4 & 5 & 7 & 12  \\ \hline
$\hat{N}_{\min}$ & 2 & 3 & 4 & 5  \\ \hline
\end{tabular}
\end{table}

First, we examine the impact of spatial correlation and $N_u$ on $\mathbb{P}_{N_u}\left(\tau\right)$ in Fig.~\ref{fig:three_lines}. We observe that our analysis with spatial correlation given by~\eqref{eq:coverage_probability} matches the simulations very well, which confirms that~\eqref{eq:coverage_probability} is an accurate approximation. Moreover, we observe that $\mathbb{P}_{N_u}\left(\tau\right)$ significantly increases when $N_u$ increases. Specifically, when $\tau=10~\textrm{dB}$, $\mathbb{P}_{N_u}\left(\tau\right)$ increases from 0.36 to 0.48 and then to 0.82 when $N_u$ increases from 1 to 2 and then to 8. Furthermore, we observe that ignoring spatial correlation leads to severe overestimation of the SINR coverage probability, which is evident when comparing the analysis ignoring spatial correlation, given by~\eqref{eq:neglecting_spatial_correlation}, with Monte Carlo simulation points. Specifically, when $\tau=14~\textrm{dB}$ and $N_u=8$, $\hat{\mathbb{P}}_{N_u}\left(\tau\right)$ is 0.83 while the simulation shows that the actual SINR coverage probability is 0.51. Indeed, spatial correlation significantly reduces the probability of successful reception at the destination UE, thus cannot be ignored.

In order to further examine the analysis error caused by ignoring spatial correlation, we investigate the minimum number of antennas needed at the destination UE to achieve a given SINR coverage probability target, $\xi$, for a given $\tau$. Mathematically, it is expressed as $N_{\min}\triangleq\min\left\{N_u:\mathbb{P}_{N_u}\left(\tau\right)>\xi\right\}$
when spatial correlation is considered. When spatial correlation is ignored, the expression is $\hat{N}_{\min}\triangleq\min\left\{N_u:\hat{\mathbb{P}}_{N_u}\left(\tau\right)>\xi\right\}$.
Table~\ref{table:number_of_antennas_needed} shows $N_{\min}$ and $\hat{N}_{\min}$ for different $\xi$ when $\tau=10~\textrm{dB}$. This table confirms the underestimation of the minimum number of antennas needed when spatial correlation is ignored. For example, when $\xi= 90\%$, if spatial correlation is ignored, the minimum number of antennas needed is 5. However, the actual minimum number of antennas needed is 12. Again, this example shows that ignoring spatial correlation is not acceptable when designing relay assisted mmWave cellular networks.

\begin{figure}[!t]
    \begin{center}
        \includegraphics[height=2.8in,width=0.95\columnwidth]{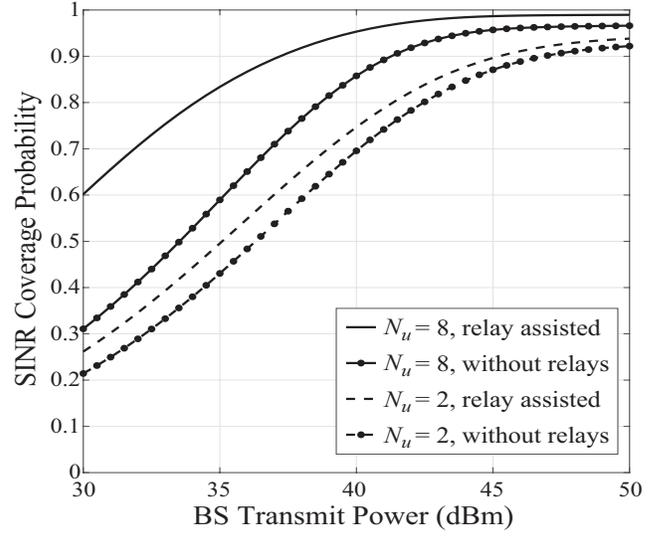}
        \caption{The SINR coverage probability, $\mathbb{P}_{N_u}\left(\tau\right)$, versus the BS transmit power, $P_b$, for different values of $N_u$, with $\tau=$10 dB.}
        \label{fig:impact_of_trans_power}
    \end{center}
\end{figure}

We now examine the impact of the transmit power at the BS, $P_b$, on $\mathbb{P}_{N_u}\left(\tau\right)$ and compare the SINR coverage probability without relays with $\mathbb{P}_{N_u}\left(\tau\right)$ in Fig.~\ref{fig:impact_of_trans_power}. We remark that the SINR coverage probability without relays is $\mathbb{P}_{N_u,bd}\left(\tau\right)$, since without relays, the relay mode does not exist. We observe that $\mathbb{P}_{N_u}\left(\tau\right)$ increases as $P_b$ increases. This is due to the fact that both $\gamma_{bd,n}$ and $\gamma_{br}$ increase as $P_b$ increases. Moreover, we observe that $\mathbb{P}_{N_u}\left(\tau\right)$ is significantly higher than $\mathbb{P}_{N_u,bd}\left(\tau\right)$. Specifically, when $\tau=10~\textrm{dB}$, $N_u=8$, and $P_b=35~\textrm{dBm}$, $\mathbb{P}_{N_u}\left(\tau\right)$ is 0.83 while $\mathbb{P}_{N_u,bd}\left(\tau\right)$ is 0.59. This confirms that introducing relays into a mmWave cellular network vastly improves the SINR coverage probability. Furthermore, we observe that the gain brought by relays, i.e., the gap between $\mathbb{P}_{N_u}\left(\tau\right)$ and $\mathbb{P}_{N_u,bd}\left(\tau\right)$, becomes larger when $N_u$ increases from 2 to 8, especially when $P_b$ is not high, e.g., $P_b=30-40~\textrm{dBm}$. This is due to the fact that when $N_u$ increases, the directional gain of relay UEs, given by $G_{U}=N_{u}$, increases. When $G_{U}$ increases, $\gamma_{br}$ and $\gamma_{rd,n}$ increase. Thus, $\mathbb{P}_{br}\left(\tau\right)$ and $\mathbb{P}_{N_u,rd}\left(\tau\right)$ increase as $N_u$ increases. Hence, based on \eqref{eq:coverage_probability},  the gain brought by relays increases as $N_u$ increases.

\begin{figure}[!t]
    \begin{center}
        \includegraphics[height=2.8in,width=0.95\columnwidth]{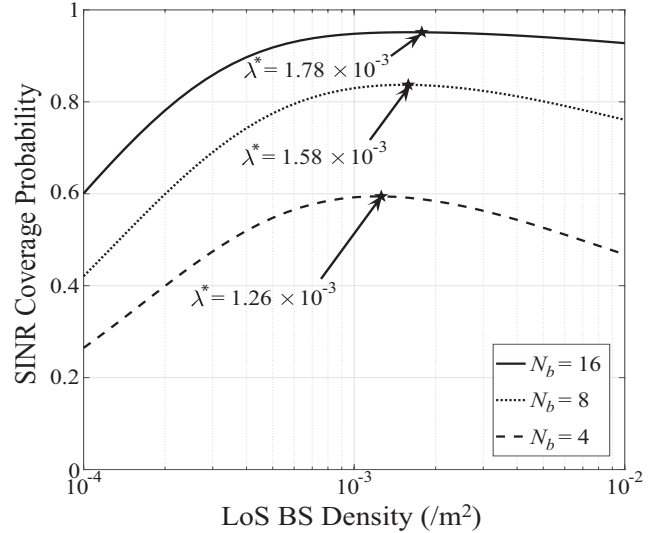}
        \caption{The SINR coverage probability, $\mathbb{P}_{N_u}\left(\tau\right)$, versus the LoS BS density, $\lambda_b$, for different values of $N_b$, with $\tau=$10 dB.}
        \label{fig:impact_of_BW_and_density}
    \end{center}
\end{figure}

Finally, we examine the impact of the LoS BS density, $\lambda_b$, and the number of antennas at the BS, $N_b$, on $\mathbb{P}_{N_u}\left(\tau\right)$ in Fig.~\ref{fig:impact_of_BW_and_density}. We observe that $\mathbb{P}_{N_u}\left(\tau\right)$ first increases then decreases as $\lambda_b$ increases. This is due to the fact that the impact of deploying more BSs is twofold. First, the associated BS is closer, which increases the desired signal power. Second, there are more interfering BSs, which increases the interference power. When $\lambda_b$ increases from $10^{-4}/\textrm{m}^2$, the first impact dominates the second. After $\lambda_b$ exceeds a certain value, which is the optimal BS density, $\lambda_b^{\ast}$, the second impact dominates the first. Thus, we highlight that $\lambda_b^{\ast}$ that maximizes $\mathbb{P}_{N_u}\left(\tau\right)$ can be numerically found with the aid of our analysis. Specifically, when $N_{b}=4$, $8$, and $16$, the optimal BS densities are $\lambda_b^{\ast}=1.26\times 10^{-3}/\textrm{m}^2$, $1.58\times 10^{-3}/\textrm{m}^2$, and $1.78\times 10^{-3}/\textrm{m}^2$, respectively. In addition, we observe that $\mathbb{P}_{N_u}\left(\tau\right)$ increases as $N_b$ increases. Specifically, when $\lambda_b=10^{-3}/\textrm{m}^2$, $\mathbb{P}_{N_u}\left(\tau\right)$ increases from 0.59 to 0.83 and then to 0.95 when $N_b$ increases from 4 to 8 and then to 16. This observation is expected because the BS beamwidth, given by $\theta_b=\frac{102 \pi}{180 N_{b}}$, decreases as $N_b$ increases and BSs with narrower beams cause less interference.

\section{Conclusion}\label{sec:conclusion}

In this paper, we proposed a new analytical framework for the relay assisted mmWave cellular network which adopts selection combining at destination UEs to avoid the misalignment problem caused by traditional directional combining. Our results showed that ignoring spatial correlation to simplify the analysis is unacceptable, since it leads to severe overestimation of the SINR coverage probability. Moreover, our results showed that introducing relays into a mmWave cellular network vastly improves the SINR coverage performance. Furthermore, we examined the impact of network parameters such as the BS density on the SINR coverage probability and found that the optimal BS density which maximizes the SINR coverage probability can be determined by using our analysis.

\section*{Acknowledgment}

This work was funded by the Australian Research Council Discovery Grant DP180104062 and the NSF Award ECCS-1642536.

\bibliographystyle{IEEEtran}
\bibliography{Simin_WCNC2020_Ref}

\begin{thebibliography}{10}
\providecommand{\url}[1]{#1}
\csname url@samestyle\endcsname
\providecommand{\newblock}{\relax}
\providecommand{\bibinfo}[2]{#2}
\providecommand{\BIBentrySTDinterwordspacing}{\spaceskip=0pt\relax}
\providecommand{\BIBentryALTinterwordstretchfactor}{4}
\providecommand{\BIBentryALTinterwordspacing}{\spaceskip=\fontdimen2\font plus
\BIBentryALTinterwordstretchfactor\fontdimen3\font minus
  \fontdimen4\font\relax}
\providecommand{\BIBforeignlanguage}[2]{{%
\expandafter\ifx\csname l@#1\endcsname\relax
\typeout{** WARNING: IEEEtran.bst: No hyphenation pattern has been}%
\typeout{** loaded for the language `#1'. Using the pattern for}%
\typeout{** the default language instead.}%
\else
\language=\csname l@#1\endcsname
\fi
#2}}
\providecommand{\BIBdecl}{\relax}
\BIBdecl

\bibitem{overview3}
S.~Rangan, T.~S. Rappaport, and E.~Erkip, ``Millimeter-wave cellular wireless
  networks: Potentials and challenges,'' \emph{Proc. IEEE}, vol. 102, no.~3,
  pp. 366--385, Mar. 2014.

\bibitem{overview4}
Z.~Pi and F.~Khan, ``An introduction to millimeter-wave mobile broadband
  systems,'' \emph{{IEEE} Commun. Mag.}, vol.~49, no.~6, pp. 101--107, June
  2011.

\bibitem{relay3}
N.~{Wei}, X.~{Lin}, and Z.~{Zhang}, ``Optimal relay probing in millimeter-wave
  cellular systems with device-to-device relaying,'' \emph{IEEE Trans. Veh.
  Technol.}, vol.~65, no.~12, pp. 10\,218--10\,222, Dec. 2016.

\bibitem{relay1}
S.~{Wu}, R.~{Atat}, N.~{Mastronarde}, and L.~{Liu}, ``Improving the coverage
  and spectral efficiency of millimeter-wave cellular networks using
  device-to-device relays,'' \emph{IEEE Trans. Commun.}, vol.~66, no.~5, pp.
  2251--2265, May 2018.

\bibitem{relay5}
N.~{Wei}, X.~{Lin}, and Z.~{Zhang}, ``Optimal relay probing in millimeter-wave
  cellular systems with device-to-device relaying,'' \emph{IEEE Trans. Veh.
  Technol.}, vol.~65, no.~12, pp. 10\,218--10\,222, Dec. 2016.

\bibitem{relay4}
G.~{Yang} and M.~{Xiao}, ``Performance analysis of millimeter-wave relaying:
  Impacts of beamwidth and self-interference,'' \emph{IEEE Trans. Commun.},
  vol.~66, no.~2, pp. 589--600, Feb. 2018.

\bibitem{relay6}
S.~{Biswas}, S.~{Vuppala}, J.~{Xue}, and T.~{Ratnarajah}, ``An analysis on
  relay assisted millimeter wave networks,'' in \emph{Proc. IEEE ICC}, Kuala
  Lumpur, Malaysia, May 2016, pp. 1--6.

\bibitem{relay7}
Y.~{Cai}, Y.~{Xu}, Q.~{Shi}, B.~{Champagne}, and L.~{Hanzo}, ``Robust joint
  hybrid transceiver design for millimeter wave full-duplex {MIMO} relay
  systems,'' \emph{IEEE Trans. Wireless Commun.}, vol.~18, no.~2, pp.
  1199--1215, Feb. 2019.

\bibitem{cover3}
T.~Bai and R.~W. Heath, ``Coverage and rate analysis for millimeter-wave
  cellular networks,'' \emph{IEEE Trans. Wireless Commun.}, vol.~14, no.~2, pp.
  1100--1114, Feb. 2015.

\bibitem{BSCoop}
C.~Skouroumounis, C.~Psomas, and I.~Krikidis, ``Low-complexity base station
  cooperation for {mmWave} heterogeneous cellular networks,'' in \emph{Proc.
  IEEE Globecom}, Washington, DC, Dec. 2016, pp. 1--6.

\bibitem{cover6}
X.~Yu, J.~Zhang, M.~Haenggi, and K.~B. Letaief, ``Coverage analysis for
  millimeter wave networks: The impact of directional antenna arrays,''
  \emph{{IEEE} J. Sel. Areas Commun.}, vol.~35, no.~7, pp. 1498--1512, July
  2017.

\bibitem{SiminICC18}
S.~Xu, N.~Yang, and S.~Yan, ``Impact of load balancing on rate coverage
  performance in millimeter wave cellular heterogeneous networks,'' in
  \emph{Proc. IEEE ICC Workshops}, Kansas City, MO, May 2018, pp. 1--6.

\bibitem{Globecom181}
Y.~M. Tsang and A.~S.~Y. Poon, ``Detecting human blockage and device movement
  in {mmWave} communication system,'' in \emph{Proc. IEEE Globecom}, Houston,
  TX, Dec. 2011, pp. 1--6.

\bibitem{cover7}
A.~Thornburg and R.~W. Heath, ``Capacity and coverage in clustered {LOS mmWave}
  ad hoc networks,'' in \emph{Proc. IEEE Globecom}, Washington, DC, Dec. 2016,
  pp. 1--6.

\bibitem{hetnet7}
H.~Elshaer, M.~N. Kulkarni, F.~Boccardi, J.~G. Andrews, and M.~Dohler,
  ``Downlink and uplink cell association with traditional macrocells and
  millimeter wave small cells,'' \emph{IEEE Trans. Wireless Commun.}, vol.~15,
  no.~9, pp. 6244--6258, Sept. 2016.

\bibitem{Book1}
F.~Baccelli and B.~Błaszczyszyn, \emph{Stochastic Geometry and Wireless
  Networks: Volume I Theory}.\hskip 1em plus 0.5em minus 0.4em\relax Hanover,
  MA: Now Publishers, 2009.

\bibitem{hetnet9}
E.~Turgut and M.~C. Gursoy, ``Coverage in heterogeneous downlink millimeter
  wave cellular networks,'' \emph{IEEE Trans. Commun.}, vol.~65, no.~10, pp.
  4463--4477, Oct. 2017.

\bibitem{blockModel2}
S.~Singh, M.~N. Kulkarni, A.~Ghosh, and J.~G. Andrews, ``Tractable model for
  rate in self-backhauled millimeter wave cellular networks,'' \emph{IEEE J.
  Sel. Areas Commun.}, vol.~33, no.~10, pp. 2196--2211, Oct. 2015.

\bibitem{GammaCDFLB}
A.~Thornburg, T.~Bai, and R.~W. Heath, ``Performance analysis of outdoor
  {mmWave} ad hoc networks,'' \emph{IEEE Trans. Signal Process.}, vol.~64,
  no.~15, pp. 4065--4079, Aug. 2016.

\bibitem{Globecom182}
M.~Haenggi, ``Diversity loss due to interference correlation,'' \emph{IEEE
  Commun. Lett.}, vol.~16, no.~10, pp. 1600--1603, Oct. 2012.

\end{thebibliography}

\end{document}